**Anti-Robust and Tonsured Statistics**
**by**
**Martin Goldberg**

*Abstract: This describes a statistical technique called "tonsuring" for exploratory data analysis in finance. Instead of rejecting "outlier" data that conflicts with the model, this strips out "inlier" data to get a clearer picture of how the market changes for larger moves.*

# I. Introduction

There are no inviolable laws of finance. Any apparent statistical regularities are just empirical tendencies, and exceptions are expected to occur. The markets are composed of many heterogeneous agents, who may sometimes choose not to act in accord with any particular model. This paper describes a data-driven statistical technique for characterizing unusual market events, that may be useful for risk management, stress test design, and portfolio optimization.

Anti-robust statistics, in the sense used in this paper, are sensitive to the tails of a distribution, but relatively unaffected by the bulk of the observations ("inliers"). This is an exploratory data analysis technique, intended to highlight infrequent features of the observed data, and does not posit any mechanism for explaining such features.

A statistical modeling procedure is considered [4] to be robust if the output is relatively insensitive to small changes in the underlying assumptions, small changes in much of the input, or large changes in some of the input ("outliers" or "anomalies"). Some robust techniques work even when almost half the data are outliers. [3,4,5] In the physical sciences these anomalies are often plausibly attributed to measurement errors; in finance there are usually separate data scrubbing procedures applied before the data enters the model. Referring to actual observed data as outliers reflects a belief that the model is correct and the data that do not fit are drawn from a separate, irrelevant, or uninteresting model, and should be ignored, or represent phenomena that the modeler does not intend to capture. Proponents of these robust models claim that the models fit the relevant data even if almost half the data do not fit. A graphical indicator that an extreme data point may not be an "outlier" (neither a mismeasurement nor a draw from a different distribution) is if the point does not sit alone, but instead is consistent with extrapolation from the bulk of the data[12].

This paper describes a set of techniques for testing how well a model fits data, or in other words, highlighting the ways in which data do not conform to the selected model[1]. This is a way of examining the question in Wigner's famous paper[2]

> "How do we know that, if we made a theory which focuses its attention on phenomena we disregard and disregards some of the phenomena now commanding our attention, that we could not build another theory which has little in common with the present one but which, nevertheless, explains just as many phenomena as the present theory?"

The paper uses correlation as the statistical measurement to illustrate these techniques. The main technique discussed is called "tonsuring." The robustification technique of trimming assigns zero weight to outliers; Winsorization downweights data according to how anomalous they are. Tonsuring is the opposite of trimming and assigns zero weight to "inliers."

An example of trimmed correlation in finance is the CDO "base correlation" curve in the Gaussian Copula model, where the attachment point is 0%, and the detachment point is the trimming point, varying from 0% to 100%. A tonsured CDO correlation curve would keep the detachment point at 100% and tonsure below the attachment point, varying from 0% (all the data) to 100%. This tonsured correlation is the correlation calculated conditional on using only relatively large changes for the input data, as defined by some distance metric.

# II. Caveats about Defining Correlation

The usual Pearson correlation

$$\rho^{P}_{X,Y} = \frac{\left(\sum_{i=1}^{n}\left(x_i - \overline{x}\right)\left(y_i - \overline{y}\right)\right)}{\sqrt{\left(\sum_{j=1}^{n}\left(x_j - \overline{x}\right)^2\right)} \bullet \sqrt{\left(\sum_{j=1}^{n}\left(y_j - \overline{y}\right)^2\right)}} \tag{1}$$

where $\overline{x}$ and $\overline{y}$ are the means,

can be calculated for any finite data set, but it can be misleading if the bivariate distribution is not elliptical, as it then depends on the size of the odd moments such as skewness. The Spearman (rank) correlation

$$\rho^{S}_{X,Y} = \frac{\left(\sum_{i=1}^{n}\left(rank(x_i) - \frac{n}{2}\right)\left(rank(y_i) - \frac{n}{2}\right)\right)}{\sqrt{\left(\sum_{j=1}^{n}\left(rank(x_i) - \frac{n}{2}\right)^2\right)} \bullet \sqrt{\left(\sum_{j=1}^{n}\left(rank(y_i) - \frac{n}{2}\right)^2\right)}} \tag{2}$$

where n/2 is the median rank,
is not sensitive to the univariate distributions, and may be more useful for highly skewed data. As an illustration, two series of Gaussian random numbers were generated with 61% Pearson correlation, and these were used to drive two Tukey *g*-and-*h* [6] processes with mean zero and unit variance. The observed Pearson correlation was calculated as a function of the two skewnesses $\gamma_1$ and $\gamma_2$, while keeping each mean at zero and each standard deviation at one. For zero skews, the driving Gaussians are recovered. Figure 1 displays the observed Pearson correlations. Note that the "true" answer is by construction 0.61. The observed rank correlation is unchanged for any skewness.

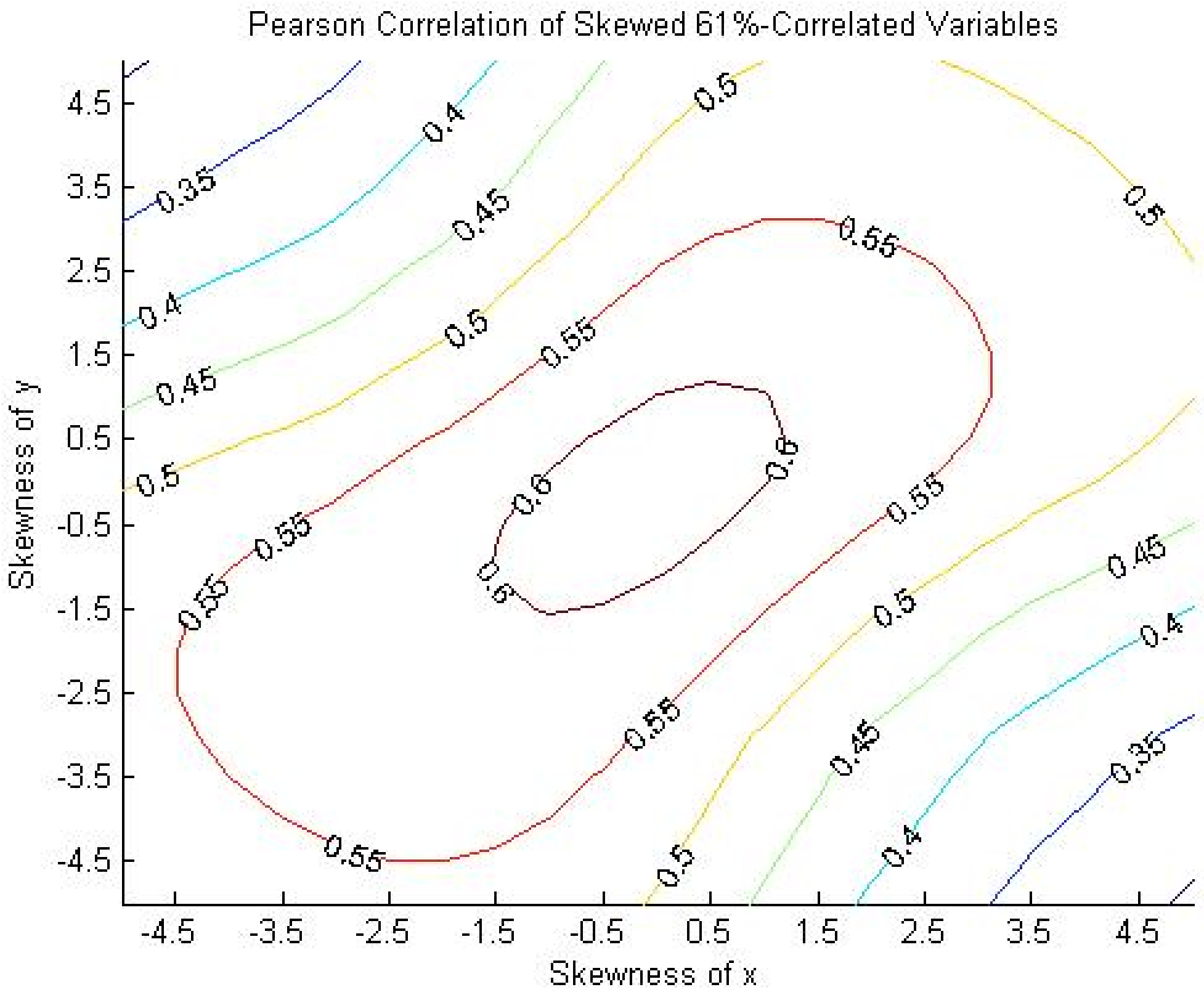

Figure 1. Measured correlation of skewed data.

## III. Definition of Tonsured Correlation

Consider a series of N sets of observations of two variables ***X*** and ***Y***. Although this is intended for a financial time series, for this paper it is assumed that

- These are independent draws from a time-invariant (stationary) distribution.
- These discrete observations come from a continuous bivariate distribution.
- All data have been scrubbed and were actually observed.
- The dataset is complete and has no issues with "missing" observations.
- Each (x,y) pair is observed contemporaneously at constant intervals – no timing issues.

To quantify which data are of more interest, some distance metric is chosen - below both the usual least-squares metric $L_2$ and the non-parametric absolute-value-of-rank $R_1$ are used. The distance $r_i$ of the i'th observation $z_i \equiv \{x_i, y_i\}$ from the centroid $\mu$ of the distribution (mean under $L_2$ and median under $R_1$) is defined as

$r_i \equiv ||z_i-\mu||$ (3)

One could use a robustified distance measure as in [3] but here only the usual $L_2$ and $R_1$ metrics are used.

We construct a distance $\delta_j$ to use as a data index and we number the data points so j=1 is closest to μ (smallest δ) and j=n is the farthest from the centroid,

$$\delta_j = \sqrt{\left((x_j - \mu_x)^2 / \sigma_x^2 + (y_j - \mu_y)^2 / \sigma_y^2\right)} \quad (4)$$

in $L_2$ or

$$\delta_j = \left|rank(x_j) - \frac{n}{2}\right| + \left|rank(y_j) - \frac{n}{2}\right| \quad (5)$$

in $R_1$.

Points closer to the center than some cutoff T, measured by $\delta_j \leq \delta_T$, are labeled "inliers" and excluded. This is called "tonsuring" after the medieval monk's practice of shaving the middle of their heads to simulate male pattern balding. The center "bald spot" is typically roughly elliptical with $L_2$ tonsuring of the data density, and typically star-shaped with $R_1$ tonsuring. The bald spot in the rank density (copula density) is typically diamond-shaped with $R_1$ tonsuring.

The ratio

[100% - (100%*(n-T)/n)] (6)

is the percentage tonsuring.

Tonsuring is not the same as center censoring, where the middle observations are all set equal to some average, but is a type of center truncation, where the middle values are discarded. Unlike [8], this technique requires no regressions and does not postulate any mechanisms – it is purely a characterization of data.

Some tonsured plots of changes in 1 year US Treasury rates versus changes in 10-year US Treasury rates are shown here. Both the probability density and copula density are separately tonsured using $R_1$ and $L_2$ tonsuring, at 50% and at 75%.

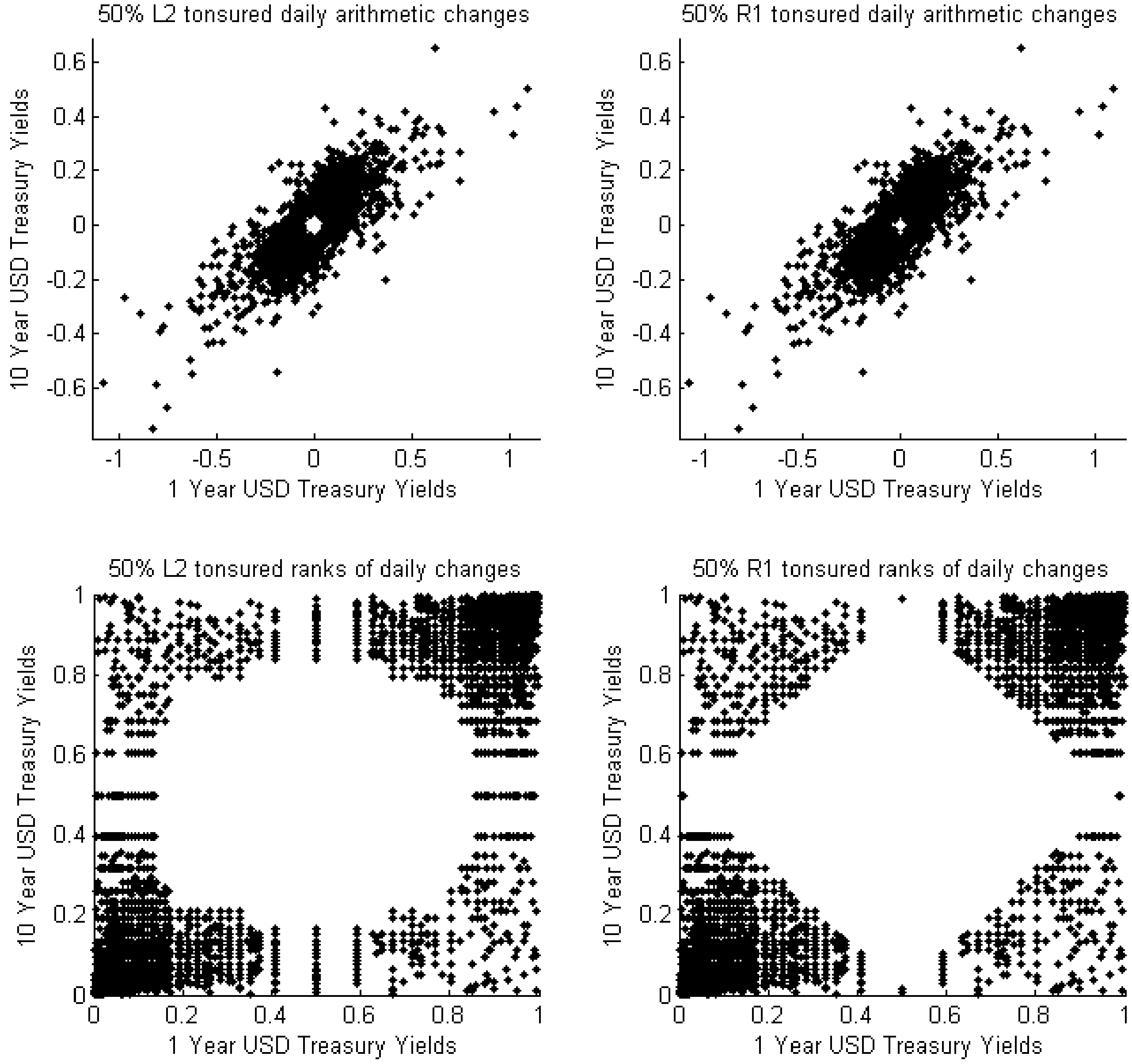

Figure 2. 50% Tonsuring
Top 2 plots are probability density, and bottom 2 plots are copula (rank) density.

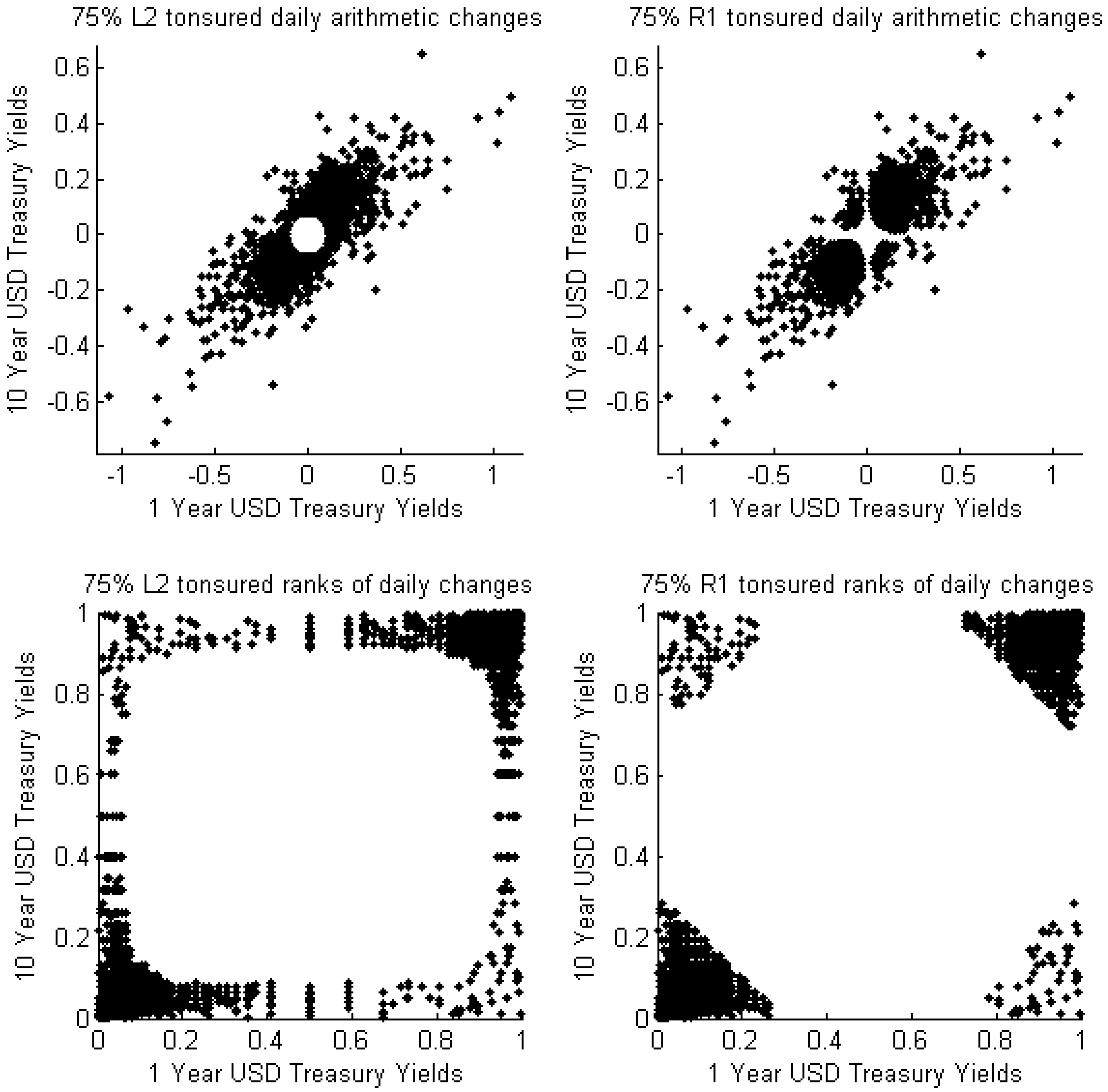

Figure 3. 75% tonsuring
Top 2 plots are probability density, and bottom 2 plots are copula (rank) density.

This paper uses three measures of association (concordance), Pearson correlation, Spearman rank correlation, and Somers $d_{BA}$.

The tonsured Pearson correlation is

$$\rho_T = \frac{\left(\sum_{i=T}^{n}\left(x_i - \overline{x_T}\right)\left(y_i - \overline{y_T}\right)\right)}{\sqrt{\left(\sum_{j=T}^{n}\left(x_j - \overline{x_T}\right)^2\right)} \bullet \sqrt{\left(\sum_{j=T}^{n}\left(y_j - \overline{y_T}\right)^2\right)}} \qquad (7)$$

Where the cutoff T is measured by $\delta_i > \delta_T$, and $\overline{x_T}$ and $\overline{y_T}$ are the means of the data remaining after discarding inliers.
Analogously, the tonsured Spearman rank correlation is calculated using the same formula (7) but substituting rank($x_i$) for $x_i$ and rank($y_i$) for $y_i$, and using (n-T)/2 for $\overline{x_T}$ and $\overline{y_T}$ .

In $R_1$ an additional analogous tonsured measure of association is a tonsured Somers's $d_{BA}$, defined by

$$C = \sum_{i=T}^{n}\sum_{j=T}^{n} 1\left(sign\left(x_i - x_j\right) = sign\left(y_i - y_j\right)\right) 1(j \neq i)$$

$$D = \sum_{i=T}^{n}\sum_{j=T}^{n} 1\left(sign\left(x_i - x_j\right) \neq sign\left(y_i - y_j\right)\right) 1(j \neq i)$$

$$T_X = \sum_{i=T}^{n}\sum_{j=T}^{n} 1\left(x_i = x_j\right) 1\left(y_i \neq y_j\right) 1(j \neq i)$$

$$T_Y = \sum_{i=T}^{n}\sum_{j=T}^{n} 1\left(y_i = y_j\right) 1\left(x_i \neq x_j\right) 1(j \neq i) \qquad (8)$$

$$S_X = \frac{C - D}{C + D + T_X}$$

$$S_Y = \frac{C - D}{C + D + T_Y}$$

$$S = d_{BA} = .5\left(S_X + S_Y\right)$$

where in the set (8) of formulas, the sums are over all $z_i$ and $z_j$ remaining after tonsuring($\delta_j > \delta_T$ and $\delta_i > \delta_T$), and the indicator functions 1(…) are of ranks recomputed after tonsuring, and not comparisons of values. Note C means concordant, D means discordant, T means tied, and S is Somers's $d_{BA}$ measure. Somers S is not expected to be numerically close to the Pearson and rank correlations - two Gaussian random variables with a Pearson correlation of ½ will have a rank correlation of about ½ but a Somers $d_{BA}$ of about ⅓.

# IV. Examples of tonsuring

Graphs of the Pearson, Spearman, and Somers correlations, as a function of tonsuring percentage, between annual S&P 500 returns and arithmetic changes in 1-year Treasury rates since 1871 (data taken from Shiller's website) are displayed as Figure 4, separately for L2 and R1 tonsuring. To verify that there is some noteworthy effect, results must be compared to the null hypothesis of multivariate normality – tonsured correlation between variables distributed multivariate Gaussian increases slightly as an artifact of the tonsuring process [7]. All 6 graphs on real data lie below the equivalent Gaussian random numbers lines.

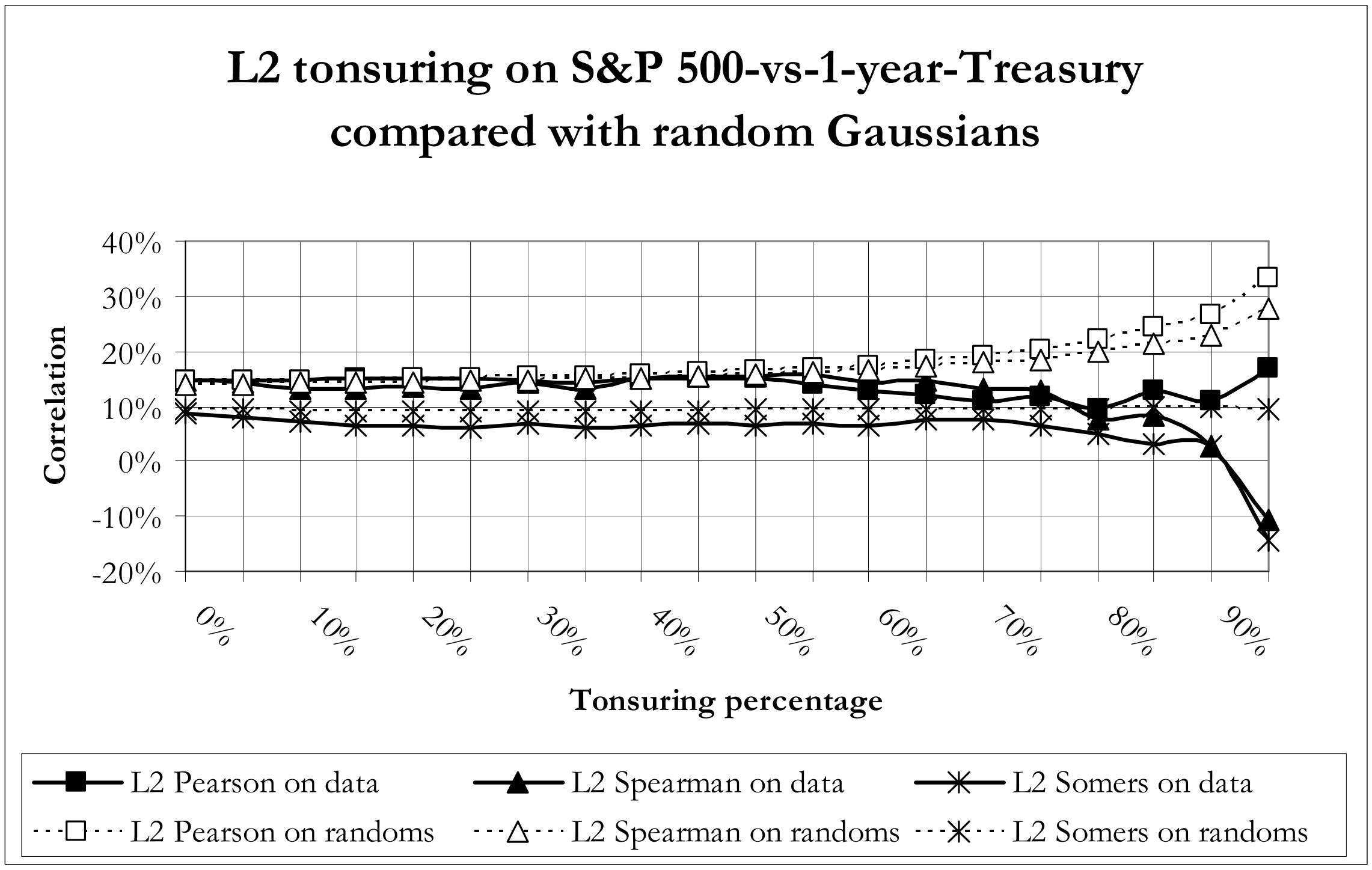

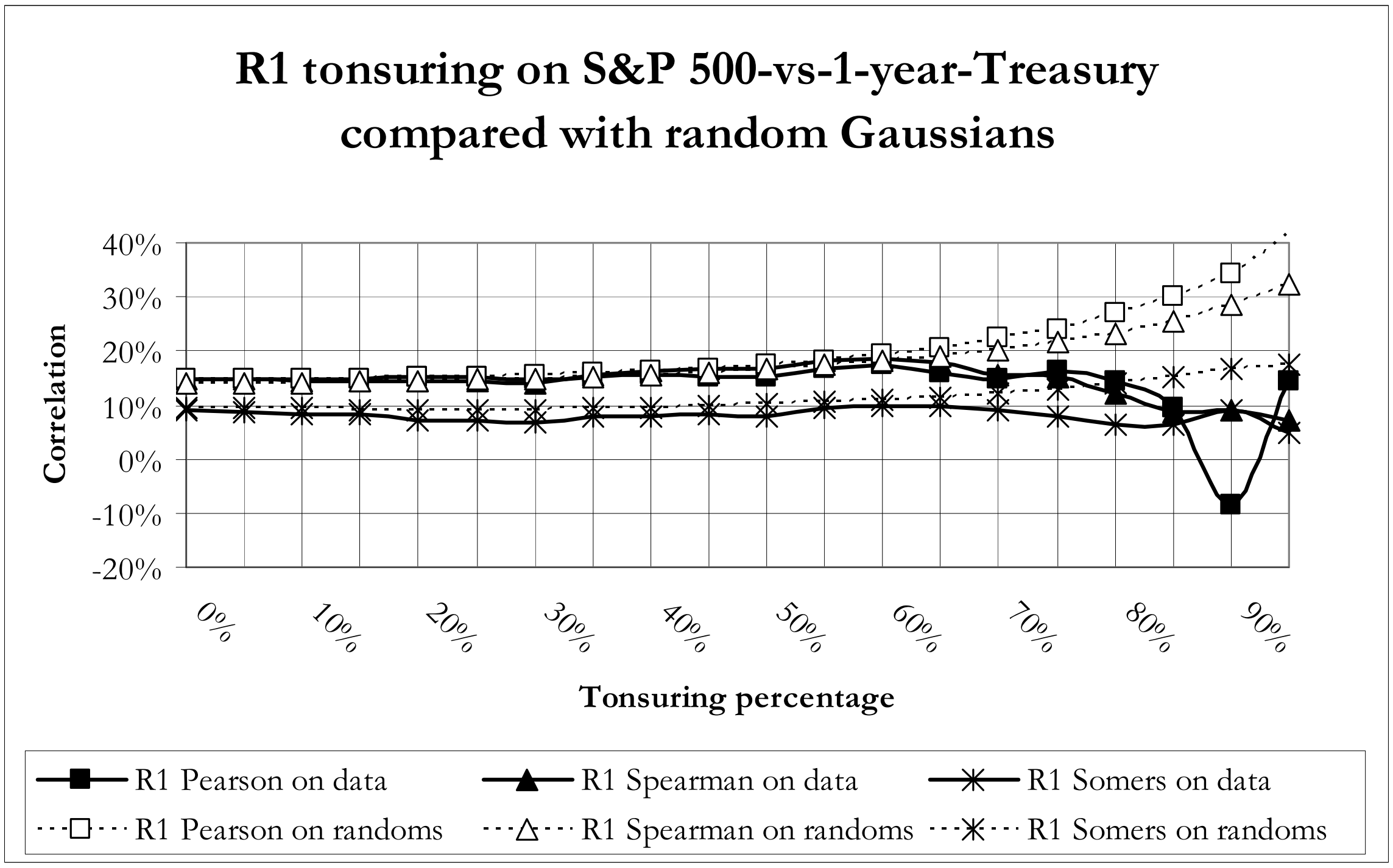

Figure 4 – Tonsured correlation between annual changes in the U.S. stock and bond markets from 1871 to 2004. The random Gaussians were generated to have the same 14.6% Pearson correlation for the full dataset as the historical data.

If there were some parametric form for the tonsured correlation, one could look for a limit as the tonsuring percentage went to 100%. This is the procedure for tail dependence, as discussed below. However, a central point of this paper is that there is in general no such parametric form for tonsured correlation that mathematically must be obeyed by the data and is suitable for extrapolation; no tonsuring equivalent to the generalized Pareto distribution of Extreme Value Theory. This is purely an explanatory data analysis, and no functional form is forced onto the data. The tonsured correlation becomes statistically meaningless after a certain level, typically when there are fewer than one or two dozen data left. This can usually be seen in a graph of $\rho$ vs tonsuring percentage as a loss of monotonicity.

To see if observed effects are more than an artifact of the tonsuring procedure, this paper will compare each pair of financial timeseries with a pair of Gaussian random numbers generated with the same full-sample Pearson correlation. It is commonly thought that, in times of stock market stress, inter-stock correlations go to either one or zero. Figure 5 examines this using L2 tonsuring of the Pearson correlation between weekly returns on BAC and on GE common stock using data from Yahoo Finance from 1986 to mid-2010. These two stocks were chosen at random purely for illustrative purposes. The correlation for the real data increases with tonsuring faster than the slight increase observed with random Gaussians generated with the same 45% full-sample Pearson correlation.

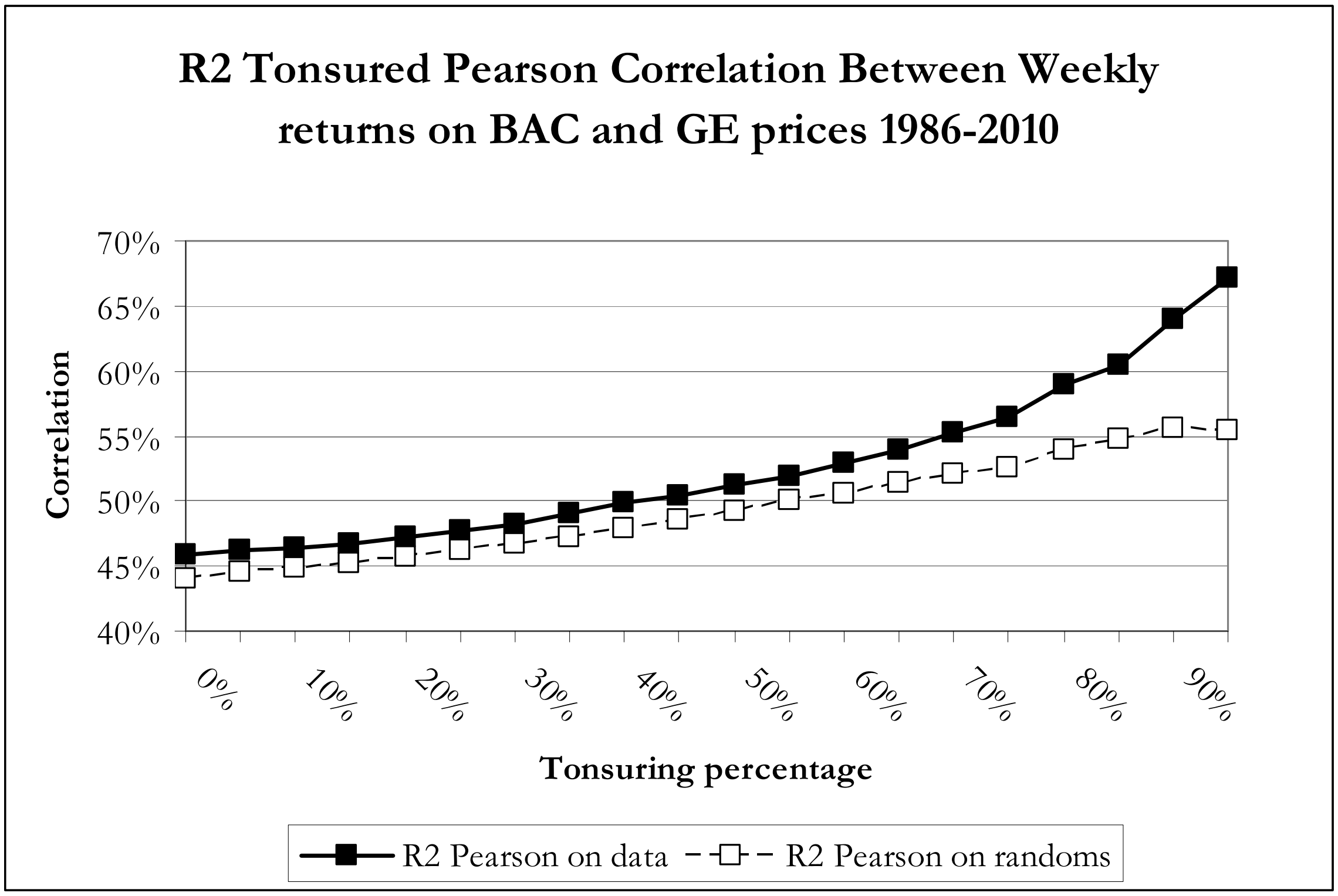

Figure 5 – R2 tonsured Pearson correlation for two stock returns

# V. Tonsuring compared to Quantile Regression and Tail Dependence

Quantile regression [8] assumes one variable is dependent on one or more explanatory variables, with the measure of dependency calibrated to optimize the parametric fit at one or more specific quantiles. The tonsuring methodology in this paper treats both variables symmetrically, and has no fitting parameters, but is purely an empirical measure.

Tail dependence is another non-parametric technique, which is complementary to tonsuring. In two dimensions, upper right tail dependence $\lambda_{ur}(u)$, as u goes to zero, is defined for any bivariate copula density c(x,y) as

$$\lim_{u\to 0}\left(\lambda_{ur}(u)\right)=\lim_{u\to 0}\left[\frac{\Pr(x>1-u,y>1-u)}{u}\right] \quad (9)$$

where changing u to 1-u and > to < in the obvious way gets the formulae for upper left (ul), bottom right (br), and bottom left (bl) tail dependencies. In words, this is a measure of whether copula density for a small square of size $u^2$ at the corner ( angle z/||z|| ) at 45˚, 135˚, 225˚, or 315˚ stays finite, or goes to zero, as the square gets smaller. The tail size u serves a similar function to the tonsuring percentage. Note that equation (9) is expressed in terms of the copula density, and not in tems of the (cumulative) copula $C(x,y)\equiv\int_0^x\int_0^y c(x,y)dxdy$.

For continuous variables, λ is bounded between zero and one, since at least zero copula density is inside the small square, and if y=x, then since Pr(x>1-u)=u by construction for any copula with no mass points, the upper limit is u/u=1. For real financial data, only certain discrete values occur due to discrete tick sizes, and there are many mass points. If some of the data points inside the $u^2$ square are observed more than once (tied ranks), λ can go above 1. There are ways to address this but they are outside the scope of this paper.

Gaussian copulas have zero tail dependence in the limit u→ 0 for all 4 of these tails – the most extreme values of one marginal never accompanies the most extreme values of the other for Gaussian copulas.[10] Tail dependence is one way of defining contagion – the tendency for local correlations in the most extreme situations to go to 1 or -1, such as is often claimed for financial crises. This tail dependence measure may not detect changes in concordance which either are more subtle than being very close to 1 or 0, or if the tails are too thin.

Since there is no universal parametric formula for real data, $\lambda_i(u)$ is plotted vs u until u is small enough that the square contains too few data.
An analogous measure, "tail insulation," or "crisis hedging," measures the tendency for one variable to go to zero when the other gets extreme. There are four of these, 0t, r0, 0b, and l0, (top, right, bottom, and left) – angles of 0º, 90º, 180º, or 270º - and the right-side one $\lambda_{r0}$ is

$$\lambda_{r0}(u)=\frac{\Pr(x>1-u,.5-u/2<y<u/2+.5)}{u} \quad (10)$$

As an example of tonsuring and tail dependence, consider daily arithmetic changes in Treasury 10-year yields from England and Japan. For ease of display, only $R_1$ tonsuring is shown. As seen in Figure 7 below, the tonsured correlation for all 3 measures is lower for the real data (heavy lines) than for the simulated bivariate normal distribution with the same untonsured Pearson correlation (dotted lines). As a consequence, in this example, if a practitioner were calculating a 99% Value-at-Risk on a portfolio sensitive to only these 2 yields, but using an approximation of a static correlation, the input correlation to get appropriate results at the 99th percentile – a 99% $R_1$ tonsured Pearson correlation of 31% - would be an input of 9%, which in this case is slightly lower than the untonsured actual measurement of 10.3%. Similarly, the asset correlations used in multivariate Merton models of default likelihood might need adjustment.

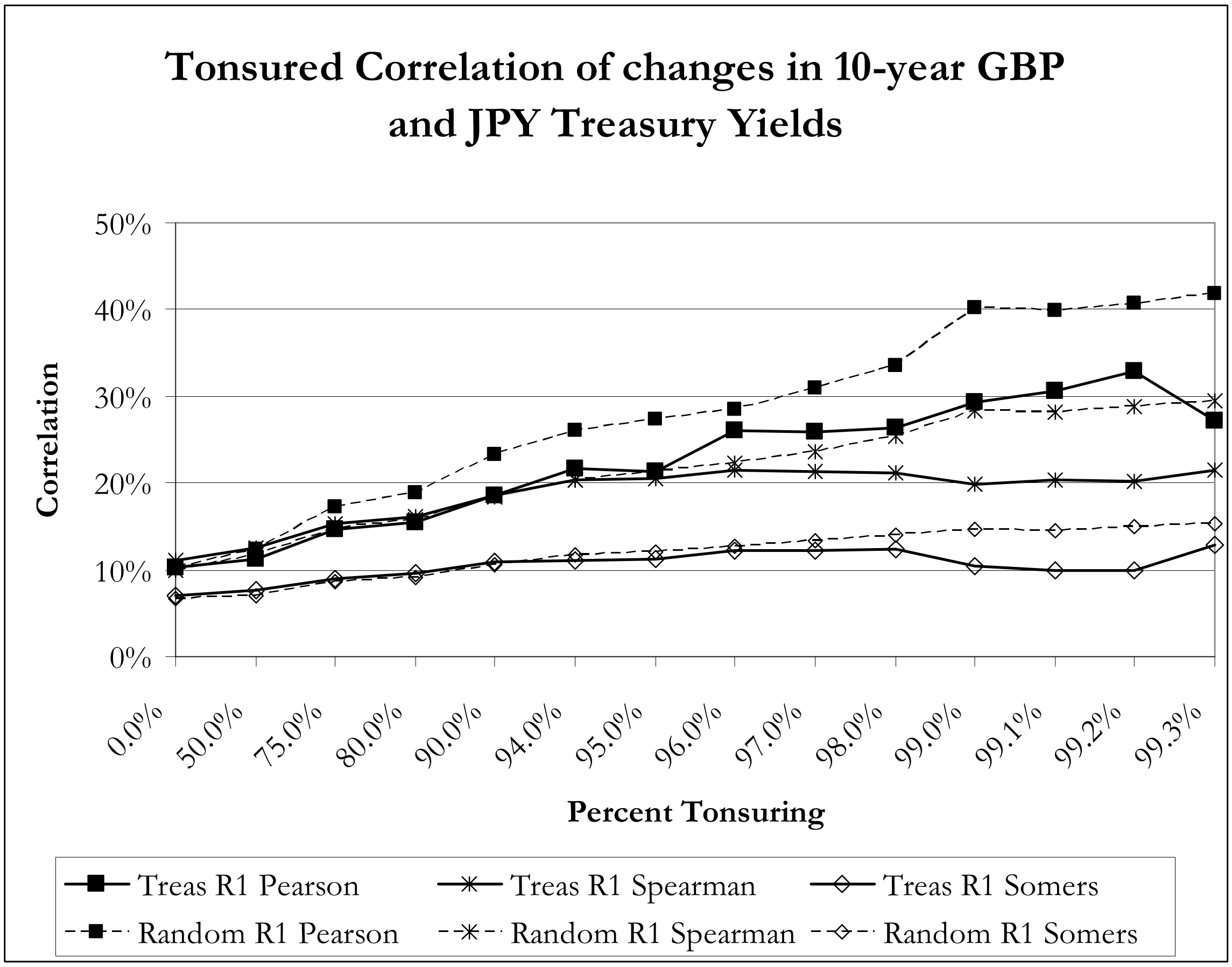

Figure 6 – R1-Tonsured correlation for daily changes in GBP and JPY 10-year Treasury yields

Figure 7 compares tail dependence of the same 10-year yield changes as Figure 6. The real data show somewhat slower decay in tail dependence than random, but go to zero or nearly zero. Only one tail, bottom right, stays above zero – a possibly insignificant tendency for large decreases in JPY rates to accompany large increases in GBP rates.

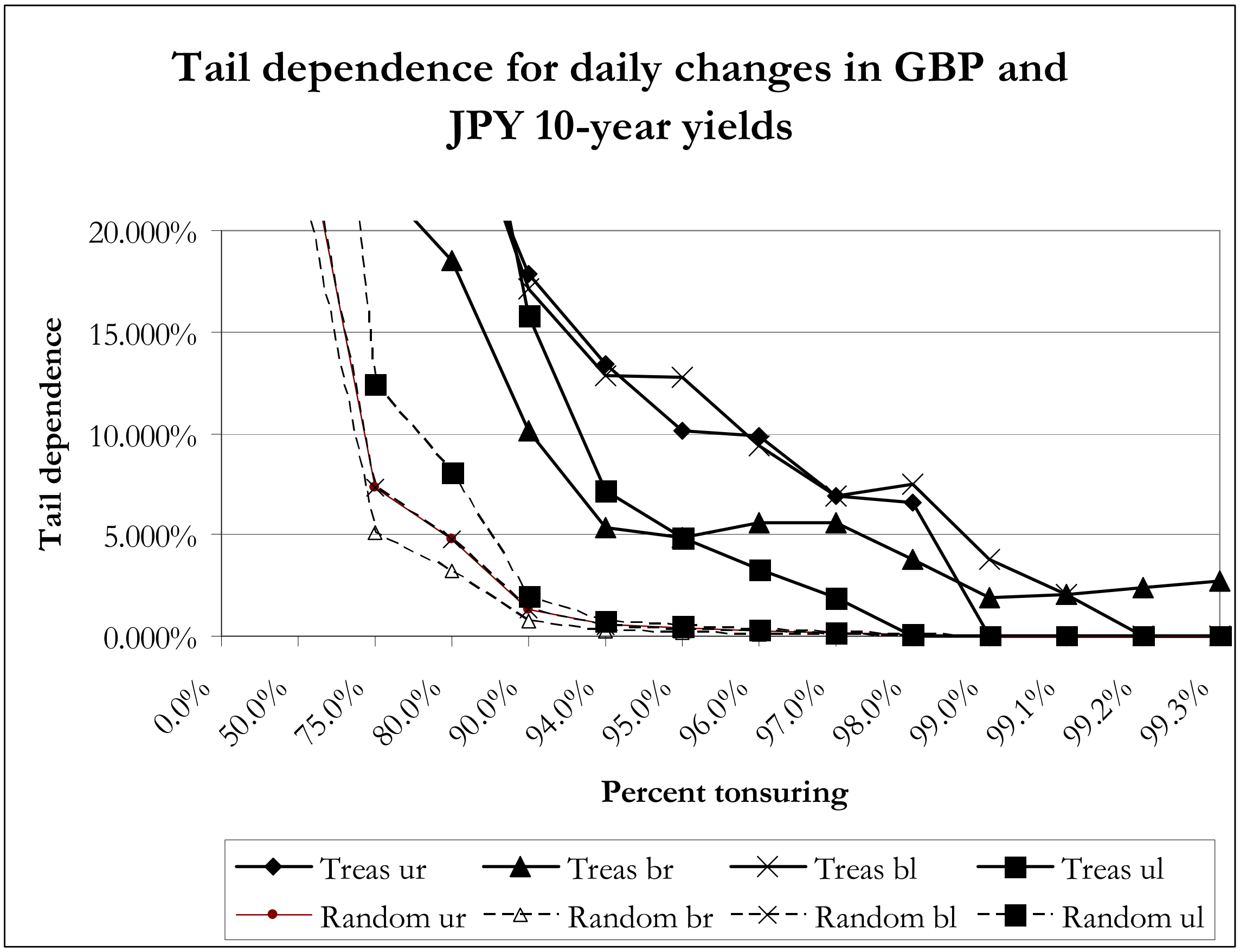

Figure 7– Tail dependence for daily changes in 10-year Treasury yields for England and Japan

Figure 8 shows tail dependence for the same weekly returns on BAC and GE stock as Figure 5. There is no meaningful difference in this case between any of the four tail-dependence graphs of the data and of random numbers, even though the tonsuring graph shows a significant increase in dependence between BAC and GE compared to Gaussian random numbers.

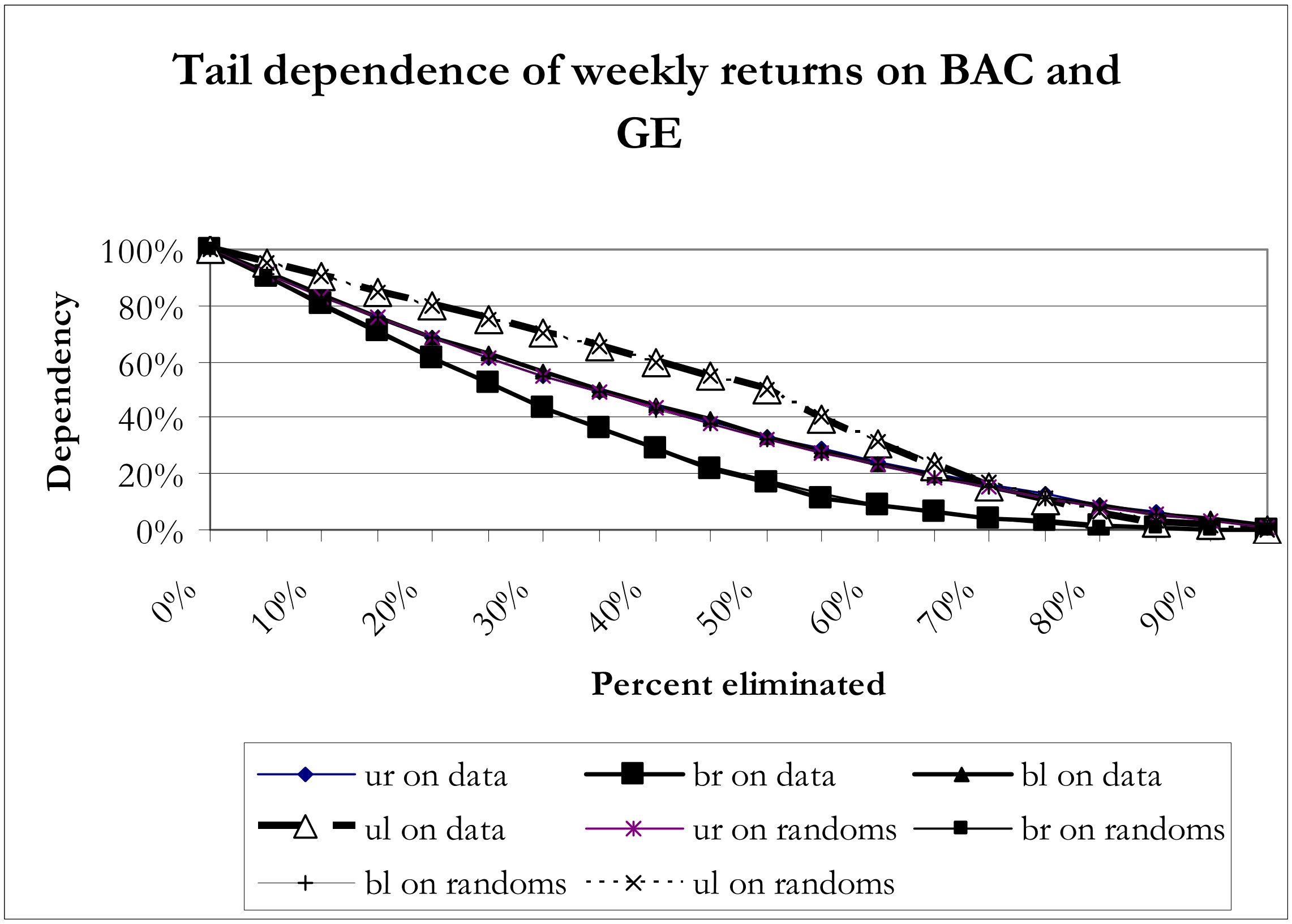

Figure 8. Tail dependence for weekly stock returns compared to correlated Gaussian random numbers.

# VI. Future Work

As noted above, this paper attempts to analyze pairs of timeseries using correlation as a weighted average over directions (angles), as a function of some distance metric, and as an average over time. There are techniques to extract more information about the angular distribution and time distribution.

A technique, described in [9] can be combined with tonsuring to get a more detailed view of the angular distribution. Describing Luo and Shevchenko's technique using copula densities, they divide the bivariate copula density into 8 regions, by dividing along the 4 lines x=.5, y=.5, x=y, and x=-y. Starting at 9:00 (x axis, x<0) and going clockwise, they number the regions 1 through 8. For an elliptical copula density, there are two sets of four equally populated octants:

$N_1=N_2=N_5=N_6$ and $N_3=N_4=N_7=N_8$. [1] An alternative measure of concordance is defined by comparing the populations of different octants.

Combining this with tonsuring k percent using the $R_1$ measure, and defining the population in each octant i after tonsuring as $N_{ki}$, and computing the averages

$$N_{ka} = \frac{N_{k1} + N_{k2} + N_{k5} + N_{k6}}{4}$$
$$N_{kb} = \frac{N_{k3} + N_{k4} + N_{k7} + N_{k8}}{4} \tag{11}$$

Then $N_{ka}/N_{kb}$ is a non-parametric tonsured measure of association, and the various ratios $N_{ki}/N_{k(a\text{ or }b)}$ measure asymmetry. Note that this is not equivalent to the tail dependence measures in equations (7) or (8).

A technique described in [11] for non-parametrically segmenting the series into different epochs with adjustable tolerance could be combined with tonsuring to get a more detailed description of time series effects in the data.

Extending tonsuring to more than 2 variables, the Principal Component Analysis of tonsured correlation matrices could be compared to the Marcenko-Pastur distribution of eigenvalues of Gaussian noise from Random Matrix Theory[13].

---

[1] This Luo-Shevchenko procedure is much more complicated and less useful using probability densities – the four lines go through the mean of each variable, the major axis, and the minor axis, which axes may not be well-defined if the probability density is too far from elliptical. The resulting probability density octants have four symmetry-related pairs rather than two sets of four.

## Other Uses of Tonsuring

Although this paper uses correlation to illustrate the concept of tonsuring, it could apply to any statistical summary measure or distribution moment where the modeler can impose some definition of metric and centroid, and attach to each data point a quantitative distance from the un-tonsured centroid. For example, one could define a tonsured CAPM beta of a stock as a function of the minimum daily percentage change in the overall market retained in the calculation. For many stocks, the tonsured beta is about the same as the untonsured beta. However, for some, as shown for BAC below, there is a noticeable effect.

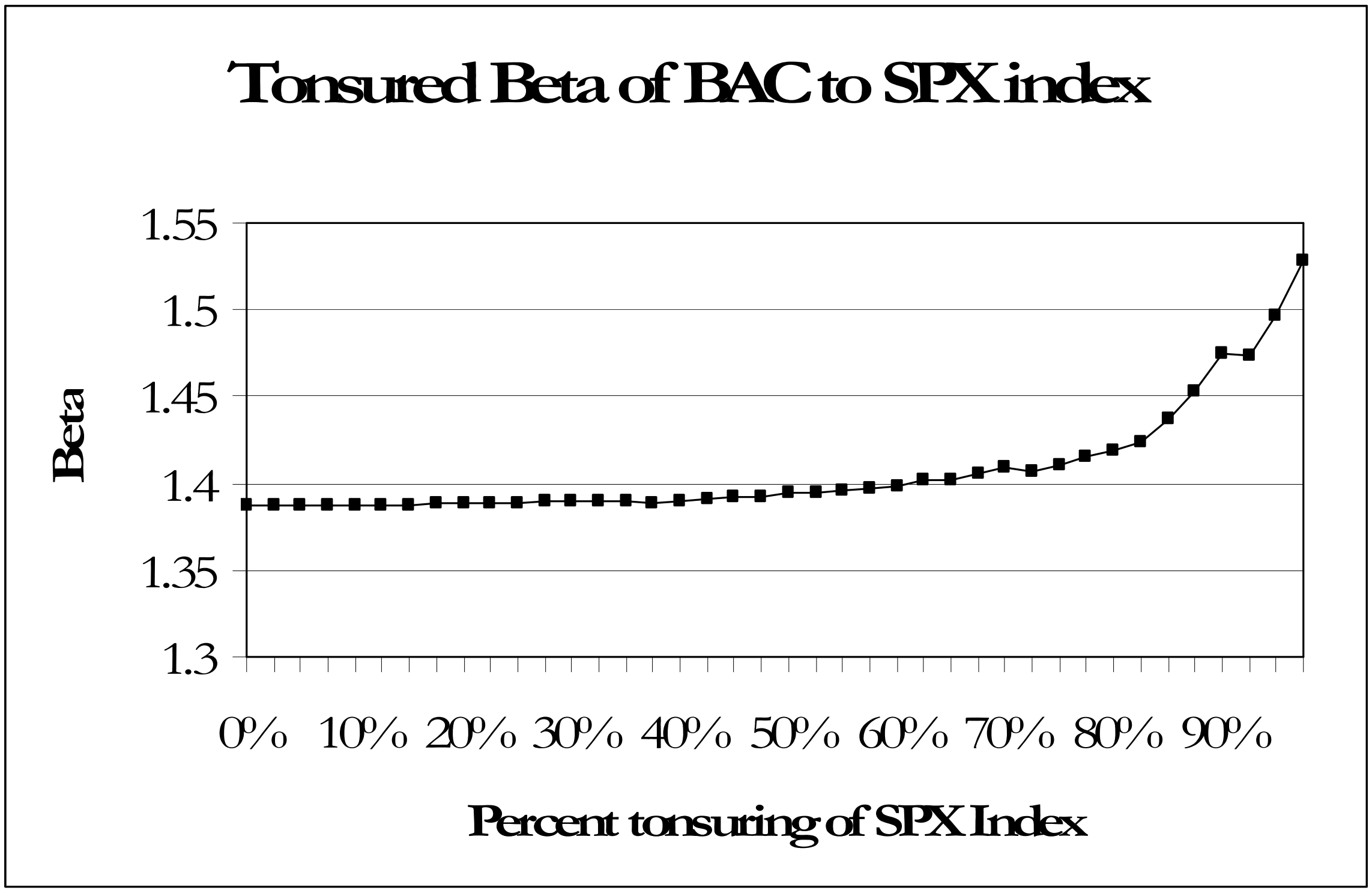

Figure 9. Tonsured CAPM beta of Bank of America stock to SPX index using daily data from Yahoo

# VII. Conclusions

By concentrating on the more extreme observations rather than dismissing them as irrelevant outliers, the tonsuring technique allows a model-free view of the more extreme data separate from the "inliers." Monte Carlo sampling from some hypothesized parametric form can be compared to the actual data in question to check for significant changes in behavior as a function of tonsuring percentage. In the examples above, the parametric form of a multivariate normal was shown to have qualitatively incorrect tonsuring behavior for the correlation between large moves for the S&P 500 vs 1-year US Treasuries, and for the correlation between two equities. A more subtle overestimate of the tonsured correlation between 10-year Treasury yield changes in England and Japan was also illustrated. In general, a significant overestimate or underestimate of tonsuring effects would suggest that a proposed parametric model would need to be modified to correctly capture tail behavior. This is a direct test of how well the model fits the data, for a model calibrated to the observed full-sample correlation.

Tonsuring can be a useful exploratory data analysis technique to highlight changes in association between pairs of financial variables as a function of size of market move, which changes may be too subtle, or the individual variables too thin-tailed, to cause asymptotic tail dependence in the sense of Equation (7). Unlike tail dependence, the tonsured correlation does not go to zero for thin-tailed data, but extrapolation to unobserved or very rarely observed extreme moves is not purely mechanical.

Tonsured correlation complements, and gives different information from, tail dependence measures. Correlations are a radial average, while tail dependences focus on one angle. The tonsuring technique can be used with several distance metrics and many definitions of correlation / dependence. For datasets with significant skewness in the univariate distribution, $R_1$ distance measures are unaffected, but $L_2$ measures should be used with caution.

Because any observed dataset is finite, there are limitations on how much data can be tonsured away while retaining statistical significance. This limitation is also present with tail dependence measures.

Tonsuring can be a useful technique to raise additional questions about the more extreme observations in a dataset, by ignoring the inliers, where not as much happens.